\documentclass{PoS}

\usepackage{epsfig}

\newcommand\half{\txf12}
\newcommand\fh[1]{{#1\over2}}
\newcommand\txf[2]{{\textstyle{#1\over#2}}}
\newcommand{\eq}{\begin{equation}}
\newcommand{\eqx}{\end{equation}}
\newcommand{\eqn}{\begin{eqnarray}}
\newcommand{\eqnx}{\end{eqnarray}}
\newcommand{\CQ}{{\cal Q}}

\PoS{PoS(LAT2005)273}

\title{Supersymmetric Yang-Mills quantum mechanics in various dimensions }

\ShortTitle{Supersymmetric Yang-Mills quantum mechanics in various dimensions }

\author{\speaker{Jacek Wosiek}\thanks{Supported by the Polish Committee for Scientific Research under
grant no. PB 1P03B 02427 (2004-2007 ).}\\
        Jagellonian University\\
        E-mail: \email{wosiek@thrisc.if.uj.edu.pl}}

\author{Massimo Campostrini \\
        INFN, Sezione di Pisa, and Universit\`{a} di Pisa\\
        E-mail: \email{Massimo.Campostrini@df.unipi.it}}

\abstract{Supersymmetric Yang-Mills quantum mechanics (SYMQM) results from
the dimensional reduction of the Yang-Mills field theory in $D$ space-time
dimensions to a
single point in the $D-1$ dimensional space. It can be also
viewed as the effective quantum mechanics of zero momentum
modes of the original theory~\cite{BSS}. These systems were first
considered in 80's~\cite{CH} as simple models with
supersymmetry~\cite{WI}. Independently, zero-volume field theories
(especially pure Yang-Mills) were employed as the starting point
of the small volume expansion, which is an important theoretical
tool complementary to early lattice
calculations~\cite{L,LM,VABA}. Later the models attracted a
new wave of interest after the hypothesis of the
equivalence, between the  $D=10, SU(\infty)$ SYMQM and a M-theory of
D0 branes~\cite{BFSS,BS,WAT,HS}.

In this talk some new results obtained for the $D=4$ system with
SU(2) gauge symmetry will be presented.}

\FullConference{XXIIIrd International Symposium on Lattice Field Theory\\
                 25-30 July 2005\\
                 Trinity College, Dublin, Ireland}

\begin{document}

The reduced quantum-mechanical Yang-Mills model is described by nine
canonically conjugate pairs of bosonic coordinates and momenta
$x^i_a(t)$, $p^i_a(t)$, $i=1,2,3$, $a=1,2,3$ and six independent
fermionic coordinates composing a Majorana spinor $\psi_a^\alpha(t)$,
$\alpha=1,...,4$, $a=1,2,3$. The Hamiltonian reads \cite{HS}
\eq
H =  {1\over 2} p_a^ip_a^i + {g^2\over 4}\epsilon_{abc}
\epsilon_{ade}x_b^i x_c^j x_d^i x_e^j + {i g \over 2}
\epsilon_{abc}\psi_a^{\dagger}\Gamma^k\psi_b x_c^k, \label{eq:Hamiltonian}
\eqx
in $D=4$, $\Gamma^k$ are the standard Dirac $\alpha^k$ matrices.

Even though the three-dimensional space was reduced to a single point,
the system still has an internal Spin(3) rotational symmetry,
inherited from the original theory, and
the residual of the local gauge
transformation.
It is also invariant under the supersymmetry transformations with
 Majorana generators
\begin{equation}
Q_{\alpha}=(\Gamma^k\psi_a)_{\alpha}p^k_a + i g
\epsilon_{abc}(\Sigma^{jk}\psi_a)_{\alpha}x^j_b x^k_c.
\label{QD4}
\end{equation}

There exists a powerful method to compute the complete spectrum and eigenstates
of polynomial hamiltonians with a "reasonably" large number of bosonic and fermionic variables
\cite{JW1,CW1,CW2,JW3}.
It  exploits the eigenbasis of the number operators associated with all individual degrees of freedom.
Beginning with the empty (fermionic and bosonic) state,
$|0\rangle=|0_F, 0_B\rangle$, we construct the basis of the physical (gauge-invariant) Hilbert
space by acting on $|0\rangle$ with gauge invariant polynomials of all creation operators.
The basis is explicitly cut by restricting
the total number of all bosonic quanta\footnote{The number of fermionic quanta is finite in reduced theories.}.
We then calculate analytically the matrix representation of the hamiltonian
and obtain numerically the spectrum and the eigenstates. The procedure is then repeated for higher cutoffs until the
results converge. Originally the analytical part of the calculation
was done in Mathematica by defining an explicit representation of the Fock states and all relevant operators. In Ref.\ \cite{CW2} this was replaced
by a fast, recursive calculation of all matrix elements. Where possible conservation laws were used.
For example, the fermionic number is conserved in $D=2$ and $D=4$ models, and the whole procedure was
carried out independently in each fermionic sector. In the four dimensional model we reduced
the problem even further by using composite creators with
fixed fermion number and angular momentum.  In the present paper, we
report yet more precise results, obtained by replacing the Mathematica
implementation of the latter algorithm with an optimized C++
program. This allowed us to obtain the energies of the first 10-20
states, in every $(F,j)$ channel, with the four digit precision.

\section{Results}
Maybe the most surprising property of these models is the coexistence of the discrete
and continuous spectrum {\em in the same range of energies}. For purely bosonic systems
the flat directions of the bosonic potential \ref{eq:Hamiltonian} are blocked by the
energy barier induced by quantum fluctuations in the transverse directions \cite{Sim}. For supersymmetric
systems, however, the above fluctuations cancel between fermions and bosons, and the continuous spectrum
appears \cite{WLN}. All these features were observed \cite{JW1} and confirmed with high accuracy
in Refs \cite{CW2}. Moreover, it was found that the continuum spectrum appears only in certain channels,
while the localized states exist for all fermion numbers and angular momenta. As a consequence, in
some channels localized  and non-localized states exist in the same energy range (see later).

     The method outlined above allowed to reach such a values of the cutoff where the discrete spectrum
and eigenstates have already converged for all practical purposes. The continuous spectrum is also well
understood.
It follows a simple scaling law which allows to extract physical information \cite{TW}.
In particular, using this scaling, a dispersion relation for the scattering states was confirmed \cite{VB2}
and obtained for more complicated fermionic sectors. Other results obtained in \cite{CW2} include
study of the restoration of supersymmetry with increasing cutoff and calculation of the Witten index.

As an illustration I would like to discuss in more detail identification of
dynamical supersymmetric multiplets in this model.

\subsection{Supersymmetry fractions}

In order to see explicitly the supersymmetric structure it is convenient
 to work with the Weyl generators
\begin{eqnarray}
{\cal Q}\dagger_{\fh1} &=& \txf12  (Q_1 - i Q_2 + Q_3 + i Q_4),
\nonumber \\
{\cal Q}^\dagger_{-\fh1} &=& \txf12 (i Q_1 + Q_2 - i Q_3 + Q_4),
 \label{WeylQ}
\end{eqnarray}
which carry definite fermionic number $F$ and
angular momentum ($m=\pm\half$). In the gauge invariant sector they
satisfy SUSY algebra
\eq
\{\CQ_m,\CQ^{\dagger}_n\}=4 H
\delta_{mn},\quad \{\CQ^{\dagger}_m,\CQ^{\dagger}_n\}=\{\CQ_m,\CQ_n\}=0.
\label{QQcom}
\eqx
It implies that supersymmetry generators can move eigenstates
of $H$ only between four channels:
\eqn
            & (F,j+1/2)  &   \nonumber           \\
(F-1,j)     &            &  (F+1,j+1) , \label{smult} \\
            & (F,j-1/2)  &  \nonumber
\eqnx
 which
form a diamond in the $F,j$ plane. Such a diamond is then nothing but a dynamical supermultiplet
with all four states  having necessarily the same eigenenergy. From now on we shall be labelling
such a  supermultiplet as $[ F,j ]$.

The algebra (\ref{QQcom}) has therefore two consequences on the spectrum: a) states from different channels
should be degenerate according to the pattern (\ref{smult}), and b) their supersymmetric images, via
$\CQ_m $ and $\CQ^{\dagger}_m $, should coincide
with others, well defined, members of a supermultiplet.

It is therefore natural to define the {\em supersymmetry fractions} as the reduced matrix elements
of Weyl generators
\[ q(F+1,j',i'|F,j,i) \equiv {1 \over 4 E_{j,i}}
   \left|{\langle F+1,j';i'\Vert \CQ^\dagger\Vert F,j ;i \rangle}\right|^2, \]
where $i$ ennumerates states in a given $(F,j)$ channel. We have found that in the limit of
exact supersymmetry and no mixing, susy fractions become simply
\begin{eqnarray}
q(F{+}1,j{+}\half|F,j) &=& q(F{+}2,j|F{+}1,j{+}\half) = j+1,
\nonumber \\
q(F{+}1,j{-}\half|F,j) &=& q(F{+}2,j|F{+}1,j{-}\half) = j,
\label{eq:Qfrac1}
\end{eqnarray}
and therefore are quite useful in identifying supermultiplets.

\subsection{Dynamical supermultiplets}

\begin{figure}[h]
\centering \leavevmode
\includegraphics[width=0.5\textwidth]{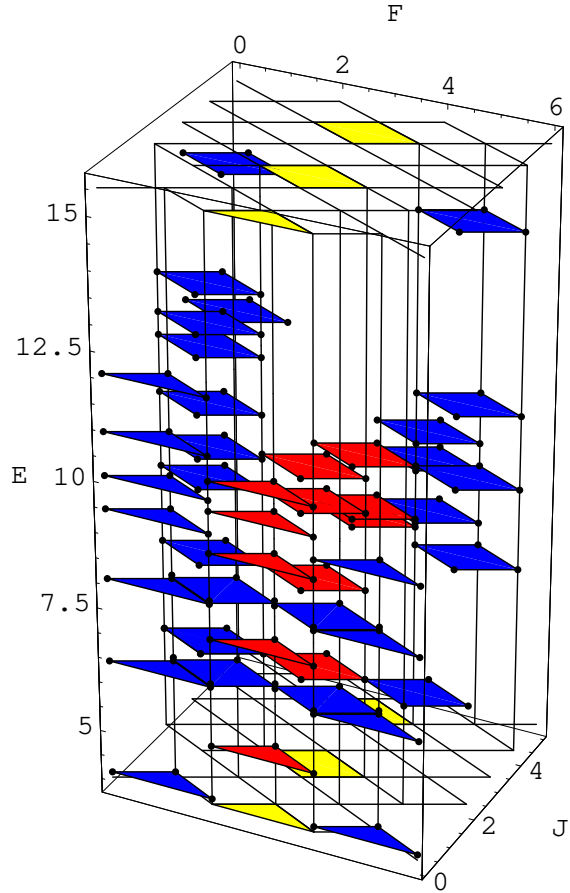}
\caption{The discrete spectrum of the $D=4$
SU(2) supersymmetric Yang-Mills quantum mechanics.  States are grouped
in supermultiplets, colored in red if the continuum spectrum is
present in the channel, in blue otherwise.}
\label{fig3D}
\end{figure}

Figure \ref{fig3D} summarizes our knowledge about the spectrum of the $D=4$, SU(2) system.
It shows a sample of eigenenergies in different channels defined by two conserved quantum numbers:
$F$ and $j$. Allowed $(F,j)$ channels are lying on the vertices of the auxilary mesh displayed
on the top and the bottom of the figure. Corresponding eigenenergies are marked by black dots.

The levels from different channels group into diamonds with the same energy.
These are the supersymmetric multiplets identified using both signatures discussed above.
 For presently available cutoffs (30-40) states within supermultiplets are degenerate
to four digits, for lower levels, and up to two digits for the highest displayed level.
Due to the particle-hole symmetry the diagram is symmetric with respect to the $F=3$ plane.
This is clearly seen for diamonds with low $E$ and $j$. For higher supermultiplets some partners
have been omitted in order not to obscure this already busy plot. Only localized states
(i.e. ones from the discrete spectrum)
 are shown. The continuous spectrum is found exclusively in the central, i.e. $F=3$, supermultiplets
and only for even $j$ (marked with the yellow color at the top and the bottom of the figure).
However in these channels the localized states exist as well, as discussed above, and again form supermultiplets.
They are marked by the red color in the figure.

The continuous spectrum  extends all the way to zero (not shown in the Figure), consequently
the SUSY vacuum must lie in the continuum with $j=0$. There are two such states with $F=2$ and $4$, hence there
are two SUSY vacua in this model. This is in correspondence with the general result for the unreduced
SU(N) field theory, that the number of vacua equals to $N$ \cite{NSVZ,RV}.

\begin{figure}[ht]
\centering
\includegraphics[width=0.4\textwidth]{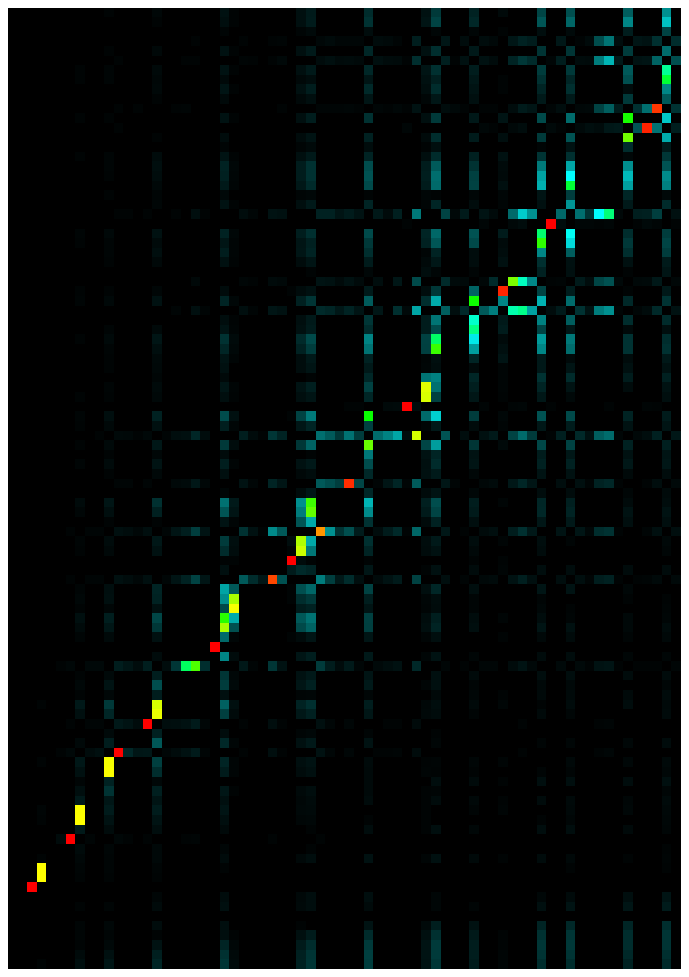}
\includegraphics[width=0.4\textwidth, angle=90]{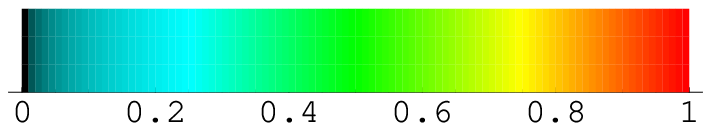}
\includegraphics[width=0.44\textwidth]{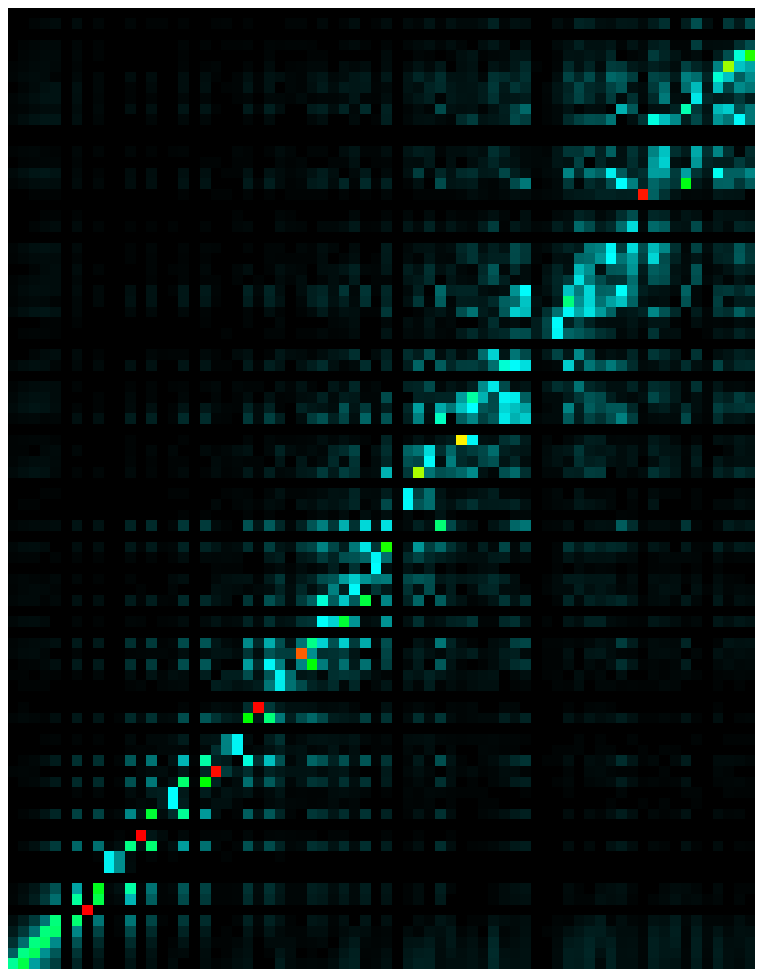}
\caption{Density plots of the supersymmetry fractions,
normalized to 1, for the channels
$(2_F,1_j)\leftrightarrow(3_F,1/2_j)$ (left: discrete spectrum), and
$(2_F,0_j)\leftrightarrow(3_F,1/2_j)$ (right: discrete and continuous spectrum).}
\label{figfr}
\end{figure}

Figure \ref{figfr} shows density plots of supersymmetry fractions,
normalized to 1, for two typical channels, i.e., transitions between
pairs of adjacent points in the $(F,j)$ plane.  Both channels show
transitions between discrete states; for
$(2_F,1_j)\leftrightarrow(3_F,1/2_j)$ we can distinguish transitions
belonging to a $[3,1]$ supermultiplets (100\% fracion, red) 
and to a $[2,1/2]$ supermultiplet (75\% fraction, yellow), see
Eq.~(\ref{eq:Qfrac1}).
The channel $(2_F,0_j)\leftrightarrow(3_F,1/2_j)$ also shows
transitions between continuum states, where the fraction is ``spread
out'' between several states and remains below 50\% (cyan and green);
this is expected, and one should perform a scaling analysis to extract
quantitative results, much in the same way as for eigenenergies. Such
an analysis is in progress.

\section{Acknowledgments}
JW thanks Gabriele Veneziano for numerous discussions and the Organizers of this Conference for their
hospitality.

\end{document}